\documentclass[twocolumn]{aastex63}
\usepackage{rotating}
\usepackage{graphicx}
\usepackage{amssymb}
\usepackage{amsmath}
\usepackage{hyperref}


\def\gtrsim{\mathrel{\hbox{\rlap{\hbox{\lower4pt\hbox{$\sim$}}}\hbox{$>$}}}}
\def\lesssim{\mathrel{\hbox{\rlap{\hbox{\lower4pt\hbox{$\sim$}}}\hbox{$<$}}}}

\begin{document}

\title{{\sl NuSTAR} observations of the Transient Galactic Black Hole Binary Candidate Swift J1858.6$-$0814: A New Sibling of V404 Cyg and V4641 Sgr? }
\author{Jeremy Hare}
\altaffiliation{NASA Postdoctoral Program Fellow}
\affiliation{Space Sciences Laboratory, 7 Gauss Way, University of California, Berkeley, CA 94720, USA}
\affiliation{NASA Goddard Space Flight Center, Greenbelt,  MD 20771, USA}
\author{John A. Tomsick}
\affiliation{Space Sciences Laboratory, 7 Gauss Way, University of California, Berkeley, CA 94720, USA}
\author{Douglas J. K. Buisson}
\affiliation{Institute of Astronomy, University of Cambridge, Madingley Road, Cambridge CB3 0HA, UK}
\affiliation{Department of Physics and Astronomy, University of Southampton, Highfield, Southampton SO17 1BJ}
\author{Ma{\"\i}ca Clavel}
\affiliation{Univ. Grenoble Alpes, CNRS, IPAG, F-38000 Grenoble, France}
\author{Poshak Gandhi}
\affiliation{Department of Physics and Astronomy, University of Southampton, Highfield, Southampton SO17 1BJ}
\author{Javier A. Garc{\'\i}a}
\affiliation{Cahill Center for Astronomy and Astrophysics, California Institute of Technology, Pasadena, CA 91125, USA}
\affiliation{Dr. Karl Remeis-Observatory and Erlangen Centre for Astroparticle Physics, Sternwartstr. 7, 96049 Bamberg, Germany}
\author{Brian W. Grefenstette}
\affiliation{Cahill Center for Astronomy and Astrophysics, California Institute of Technology, Pasadena, CA 91125, USA}
\author{Dominic J. Walton}
\affiliation{Institute of Astronomy, University of Cambridge, Madingley Road, Cambridge CB3 0HA, UK}
\author{Yanjun Xu}
\affiliation{Cahill Center for Astronomy and Astrophysics, California Institute of Technology, Pasadena, CA 91125, USA}

\email{jeremy.hare@nasa.gov}

\begin{abstract}
Swift J1858.6$-$0814 was discovered by {\sl Swift-}BAT on October 25, 2018. Here we report on the first follow-up {\sl NuSTAR} observation of the source, which shows variability spanning two orders of magnitude in count rate on timescales of $\sim$10-100 s. The power-spectrum of the source does not show any quasi-periodic oscillations or periodicity, but has a large fractional rms amplitude of 147\%$\pm3\%$, exhibiting a number of large flares throughout the observation. The hardness ratio (defined as $R_{10-79 \rm keV}/R_{3-10 \rm keV}$) of the flares tends to be soft, while the source spans a range of hardness ratios during non-flaring periods. The X-ray spectrum of the source shows strong reflection features, which become more narrow and peaked during the non-flaring intervals. We fit an absorbed  relativistic reflection model to the source spectra to place physical constraints on the system. Most notably, we find that the source exhibits a large and varying intrinsic absorbing column density ($N_{\rm H}=1.4-4.2\times10^{23}$ cm$^{-2}$). This large intrinsic absorption is further supported by the energy spectra extracted from two flares observed simultaneously by {\sl NuSTAR} and {\sl NICER}. We find that the inner accretion disk of the source has a low inclination, $i<29^{\circ}$ ( 3$\sigma$ upper-limit), while the iron abundance in the disk is close to solar, $A_{\rm Fe}=1.0\pm0.3$. We set a 90\% confidence upper limit on the inner radius of the accretion disk of $r_{\rm in}<8 \ r_{\rm ISCO}$, and, by fixing $r_{\rm in}$ to be at $r_{\rm ISCO}$, a 90\% confidence lower-limit on the spin of the black hole of $a^{*}>0.0$. Lastly, we compare the properties of Swift J1858.6$-$0814  to those of V404 Cygni and V4641 Sgr, which both show rapid flaring and a strong and variable absorption. 
\end{abstract}

\section{Introduction}

Since its launch in 2012, the {\sl Nuclear Spectroscopic Telescope Array} ({\sl NuSTAR}; \citealt{2013ApJ...770..103H}) has played a pivotal role in studying, and sometimes identifying, the nature of Galactic hard X-ray transients discovered by e.g., {\sl Swift}-BAT, {\sl INTEGRAL}, or {\sl MAXI}. Among some of the most interesting sources include a new magnetar \citep{2013ApJ...770L..23M}, several super-giant fast X-ray transients (SFXTs; e.g., \citealt{2015MNRAS.447.2274B,2019arXiv190303210F,2019ApJ...878...15H}), and numerous low-mass X-ray binaries (LMXBs) hosting either a neutron star (e.g., \citealt{2018MNRAS.474.4432J,2018ApJ...853..157H}) or a black hole (BH; e.g., \citealt{2017ApJ...851..103X,2019MNRAS.485.3064B}). Following the detection and identification of a new BH transient candidate, {\sl NuSTAR} has also helped to constrain the physical parameters of these systems through spectral fitting, using a combination of reflection models and {\sl NuSTAR}'s unprecedented sensitivity above 10 keV (see e.g., \citealt{2018ApJ...852L..34X,2018ApJ...865...18X,2019MNRAS.490.1350B}).

As a BH transient undergoes an outburst it typically evolves through several spectral states, showing relatively slow variability on kilosecond to day long timescales (see e.g., \citealt{2006ARA&A..44...49R,2016ASSL..440...61B}). These outbursts usually start in the hard state, in which the X-ray spectrum is dominated by a hard power-law component. The source then transitions into the soft state, where the X-ray spectrum becomes dominated by the hot thermal emission from the accretion disk. Finally, the BH transient returns back to the hard state at the end of the outburst. While in the soft state, the accretion flow is expected to reach the inner-most stable circular orbit (ISCO) of the BH, whose radius ($r_{\rm ISCO}$) depends on the spin of the BH. Modeling the X-ray spectra during these spectral states with relativistic reflection models allows for constraints to be placed on the BH's spin.

While the majority of BH transients generally follow the standard progression through the spectral states outlined in the previous paragraph, there are a few outliers, such as V404 Cyg and V4641 Sgr. These systems exhibit large amplitude flares, with X-ray count rates rising by factors of 10$^{2}-10^{3}$ on time scales of seconds to minutes, and reaching Eddington or even super-Eddington luminosities \citep{2000ApJ...528L..93W,2002A&A...391.1013R,2017MNRAS.471.1797M,2017ApJ...839..110W,2017NatAs...1..859G}. Accompanying these flares are significant changes in the shape of the X-ray spectra, including variations of the intrinsic absorbing column density, photon index, and reflection strength  (see e.g., \citealt{2000ApJ...528L..93W,2017MNRAS.471.1797M,2017ApJ...839..110W}). The large scale flaring behavior and rapid changes in the X-ray spectrum of these sources makes it difficult to characterize their spectral state, however relativistic reflection modeling can still place constraints on their physical parameters (see e.g., \citealt{2017ApJ...839..110W}).

Swift J1858.6-0814 (J1858, hereafter), discovered as a Galactic ($l$=26.395$^{\circ}$, $b$=$-$5.351$^{\circ}$) X-ray transient by {\sl Swift-}BAT on October 25, 2018, is a new  BH candidate exhibiting similar characteristics to V404 Cyg and V4641 Sgr \citep{2018ATel12151....1K,2018ATel12158....1L}. The source was subsequently followed-up by {\sl NuSTAR} and {\sl NICER}. The {\sl NICER} data showed that the source exhibited large amplitude flares on timescales as short as $\sim10$ s, the largest of which had a peak count rate of $\sim1000$ cts s$^{-1}$ and lasted roughly 15 s \citep{2018ATel12158....1L}. The {\sl NICER} spectra were divided into high ($>100$ cts s$^{-1}$), moderate ($20-100$ cts s$^{-1}$), and low ($<20$ cts s$^{-1}$) intensity intervals, and fit with an absorbed thermal disk plus power-law model. The best-fit models found fairly low disk temperatures of $\sim$0.2-0.3 keV, which are rather typical for BHs in the hard state (see e.g., \citealt{2010MNRAS.402..836R,2013ApJ...769...16R}), while the power-law component was found to be very hard, $\Gamma\sim1$  \citep{2018ATel12158....1L}. Additionally, the {\sl NICER} spectra showed both Fe L and K reflection features.

J1858's longer wavelength counterpart was first detected as a variable UV source by {\sl Swift}-UVOT, and it was found that the source was coincident with a previously detected UKIDSS and Pan-STARRs source \citep{2018ATel12160....1K}. Additional optical follow-up found that the source had brightened by $\sim2.5$ magnitudes over the source's cataloged Pan-STARRs $r'$ magnitude \citep{2018ATel12164....1V}. The source has also been detected in radio by AMI-LA and appears to be variable, having a flux density of 300-600 $\mu$Jy at 15.5 GHz \citep{2018ATel12184....1B}. 

\begin{figure*}
\centering
\includegraphics[trim={0 0 0 0},width=18.0cm]{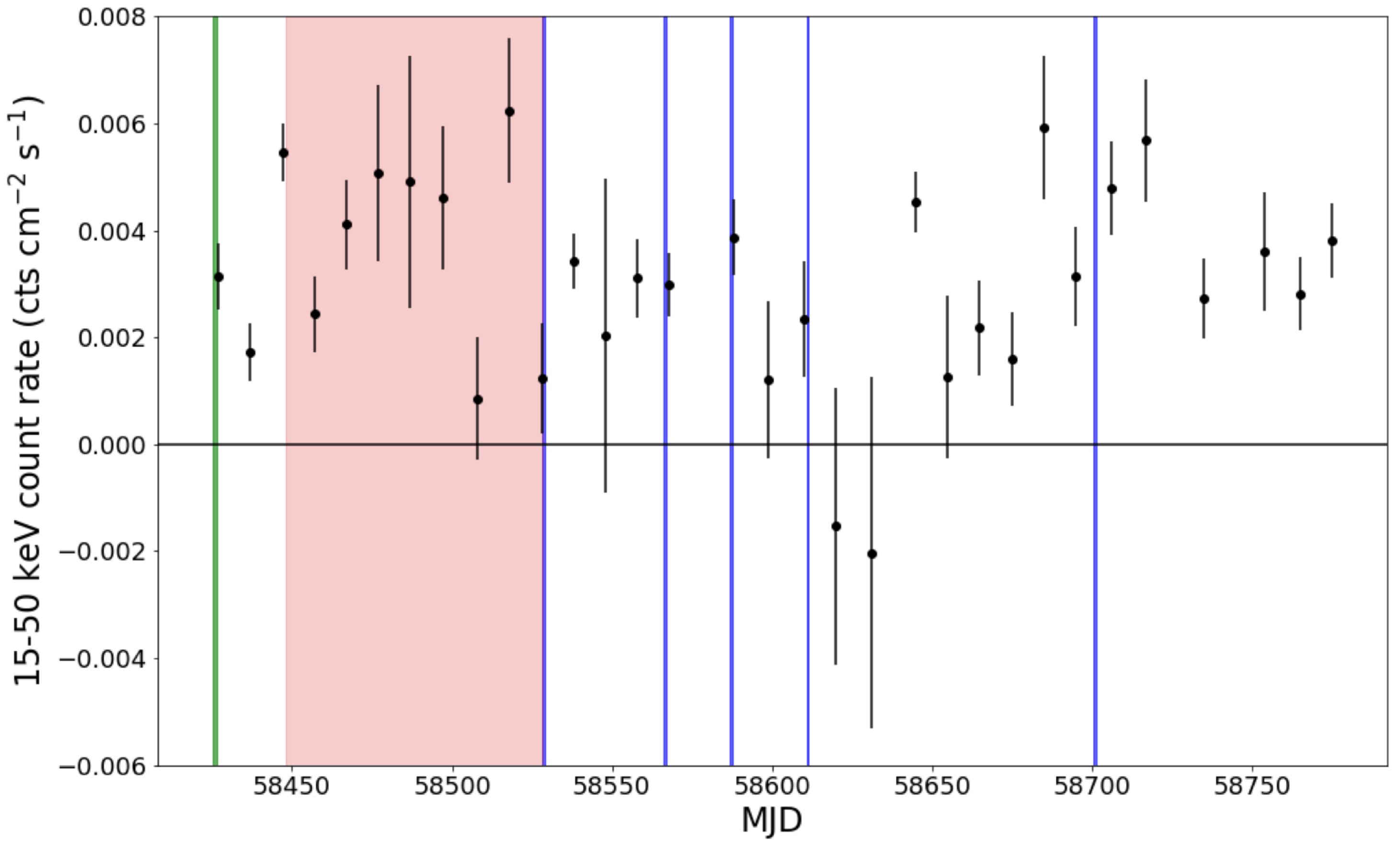}
\caption{{\sl Swift}-BAT light curve of Swift J1858.6$-$0814 (with 10-day averaged time bins) up to October 28, 2019 (to date the outburst is still ongoing). The thin vertical green and blue lines show times of the six {\sl NuSTAR} observations of the source. The green line marks the observation reported here. The wide pink band shows the time in which the source was Sun constrained for {\sl NuSTAR}.
\label{BAT_nu}
}
\end{figure*}

The source's outburst is still ongoing to date\footnote{See \url{https://swift.gsfc.nasa.gov/results/transients/weak/SWIFTJ1858.6-0814/}} (i.e., October 2019), having a mean {\sl Swift-BAT} \citep{2013ApJS..209...14K} flux of $\approx14$ mCrab (see Figure \ref{BAT_nu}). Further, the source has now been observed six times by {\sl NuSTAR} (see Figure \ref{BAT_nu}). Here we report the results of the analysis of the first of these {\sl NuSTAR} observations. In Section \ref{obs_dr}, we describe the details of the {\sl NuSTAR} and {\sl NICER} observations and data reduction, then in Section \ref{res} we discuss the data analysis and results. Next, in Section \ref{diss}, we discuss the physical parameters of this system and compare them to other similar systems, namely V404 Cyg and V4641 Sgr. Lastly, we summarize our findings in Section \ref{summ}.

\section{Observations and Data Reduction}
\label{obs_dr}

\subsection{{\sl NuSTAR}}
J1858 was observed with the {\sl Nuclear Spectroscopic Telescope Array} ({\sl NuSTAR}; \citealt{2013ApJ...770..103H}) on 2018, November 03 (MJD 58425.28, obsID 80401317002) for $\sim52$ ks after correcting for deadtime. The data were reduced using the {\sl NuSTAR} Data Analysis Software (NuSTARDAS) package version 1.8.0 and the 20181022 version of the Calibration database (CALDB). First, the photon arrival times were corrected to the solar system barycenter using {\tt nupipeline}, which also includes a clock correction from the CALDB to account for {\sl NuSTAR}'s clock-drift\footnote{See \url{http://www.srl.caltech.edu/NuSTAR_Public/NuSTAROperationSite/clockfile.php}}. Then, the energy spectra and light curves of the source were extracted from both the FPMA and FPMB detectors using a $r=90''$ circle centered on the source. The corresponding background spectra and light curves were extracted from a source free circular region ($r\approx50''$) on the same detector chip as the source.  Prior to fitting, the energy spectra were grouped to have a signal-to-noise ratio of at least five in each energy bin. We fit all X-ray energy spectra with XSPEC version 12.10.1 \citep{1996ASPC..101...17A}, accounting for interstellar absorption with v2.3.2 of the Tuebingen-Boulder ISM absorption model, {\tt tbnew}\footnote{See \url{http://pulsar.sternwarte.uni-erlangen.de/wilms/research/tbabs/}}, with solar abundances adopted from \cite{2000ApJ...542..914W}.  All uncertainties in this paper are reported at the 90\% confidence level unless otherwise noted. 

\subsection{{\sl NICER}}
\label{nice_sec}
The {\sl Neutron Star Interior Composition Explorer} ({\sl NICER}; \citealt{2012SPIE.8443E..13G}) observed J1858 twice simultaneously with the {\sl NuSTAR} observation reported here. The two {\sl NICER} observations had exposure times of $\sim 5.7$ ks and $5.2$ ks, for obsIDs 1200400103 and 1200400104, respectively. The {\sl NICER} data were reduced following the standard processing and filtering procedures using the NICERDAS (V005) software package and the {\tt xti20190520} version of the CALDB. We also excluded data from two detectors that are known to exhibit increased detector noise\footnote{See \url{https://heasarc.gsfc.nasa.gov/docs/nicer/data_analysis/nicer_analysis_tips.html} for additional details}. We use the latest response functions (i.e., {\tt nixtiaveonaxis20170601v002.arf} and {\tt nixtiref20170601v001.rmf}) for the spectral analyses of the {\sl NICER} data performed in this paper. To minimize the effects of residuals that still remain in the {\sl NICER} response functions, and to further minimize the effects of detector noise, we restrict our analysis to the 0.5-7 keV energy range\footnote{See, for example, \url{https://heasarc.gsfc.nasa.gov/docs/nicer/data_analysis/nicer_analysis_tips.html}}.

The {\sl NICER} data are primarily used in this paper to constrain the soft part of J1858's X-ray spectrum.  Since J1858 shows significant spectral evolution during its flares, we only used bright flares simultaneously observed by {\sl NICER} and {\sl NuSTAR}. During the first observation, {\sl NICER} unfortunately observed J1858 while the source was occulted by the Earth for {\sl NuSTAR}, so there is only $\sim1400$ s of strictly simultaneous data, none of which contains any particularly bright flares. Therefore we do not use any of the data from the first observation. However, the second {\sl NICER} observation overlapped with the {\sl NuSTAR} observation for $\sim4.5$ ks, and caught two of the brightest flares observed by {\sl NuSTAR}, occurring $\sim570$ s apart (see the inset in Figure. \ref{lc_100s}). These flares are referred to as flare 1 and flare 2, hereafter, with flare 1 occurring first, and flare 2 being brighter.

\section{Results}
\label{res}

\subsection{Variability and Timing}
\label{v_and_tim}
\begin{figure*}
\centering
\includegraphics[trim={0 0 0 0},width=18.0cm]{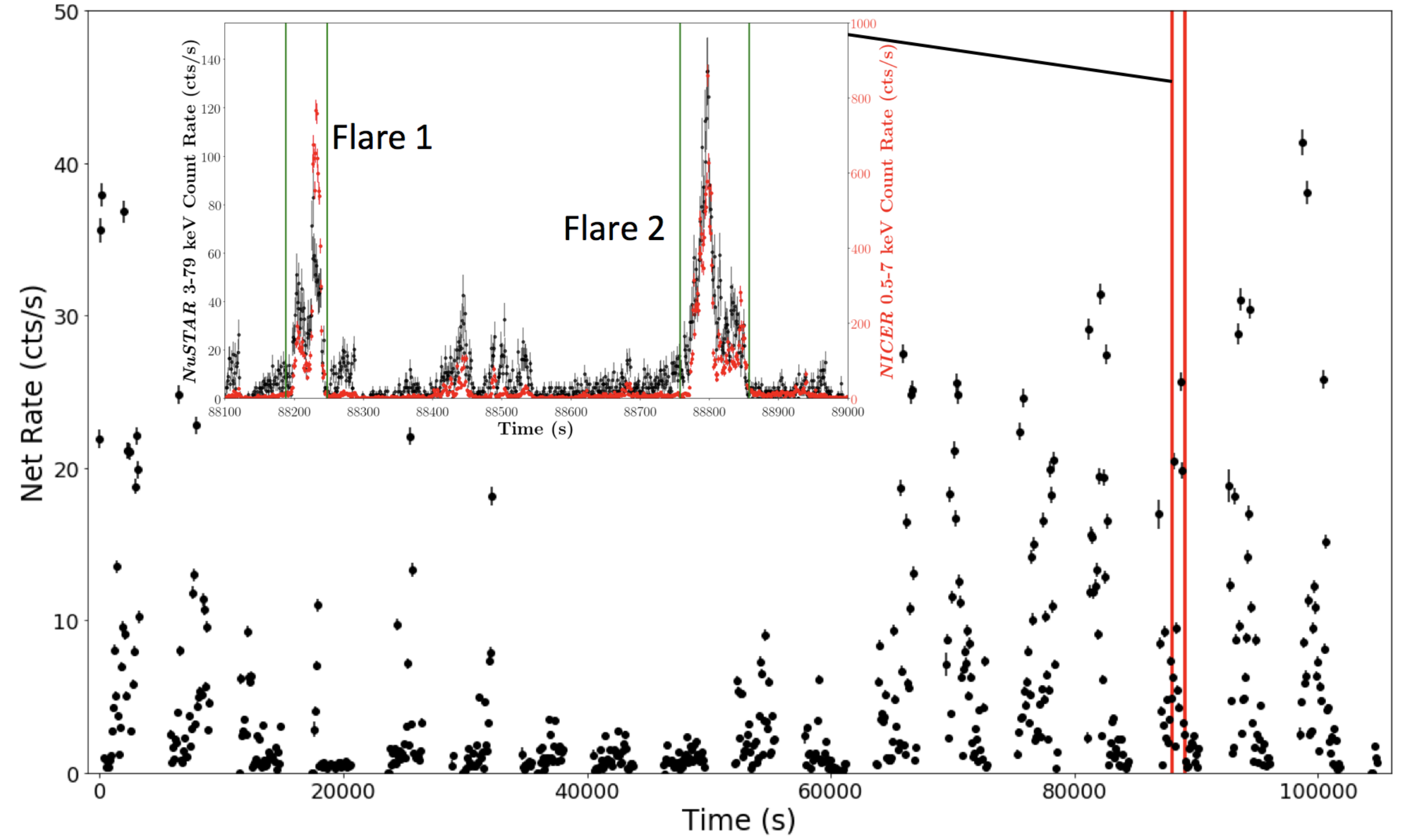}
\caption{The FPMA light curve of Swift J1858 with 100 s time bins. We show only the FPMA light curve for clarity. The inset shows a zoom in with 1 s time bins over a $\sim1$ ks span (denoted by the vertical red lines)  that includes the two large flares that occurred during the simultaneous {\sl NuSTAR} and {\sl NICER} observations (see Sections \ref{nice_sec} and \ref{nu_nice_flare}). {\sl NICER}'s count rates in the 0.5-7 keV energy range, shown as red points in the inset, are roughly a factor of 10 larger than {\sl NuSTAR}'s count rates. The green lines in the inset show the time intervals in which each flare's energy spectra were extracted (see Section \ref{nu_nice_flare}).
\label{lc_100s}
}
\end{figure*}

\begin{figure}
\centering
\includegraphics[trim={0 0 0 0},width=8.7cm]{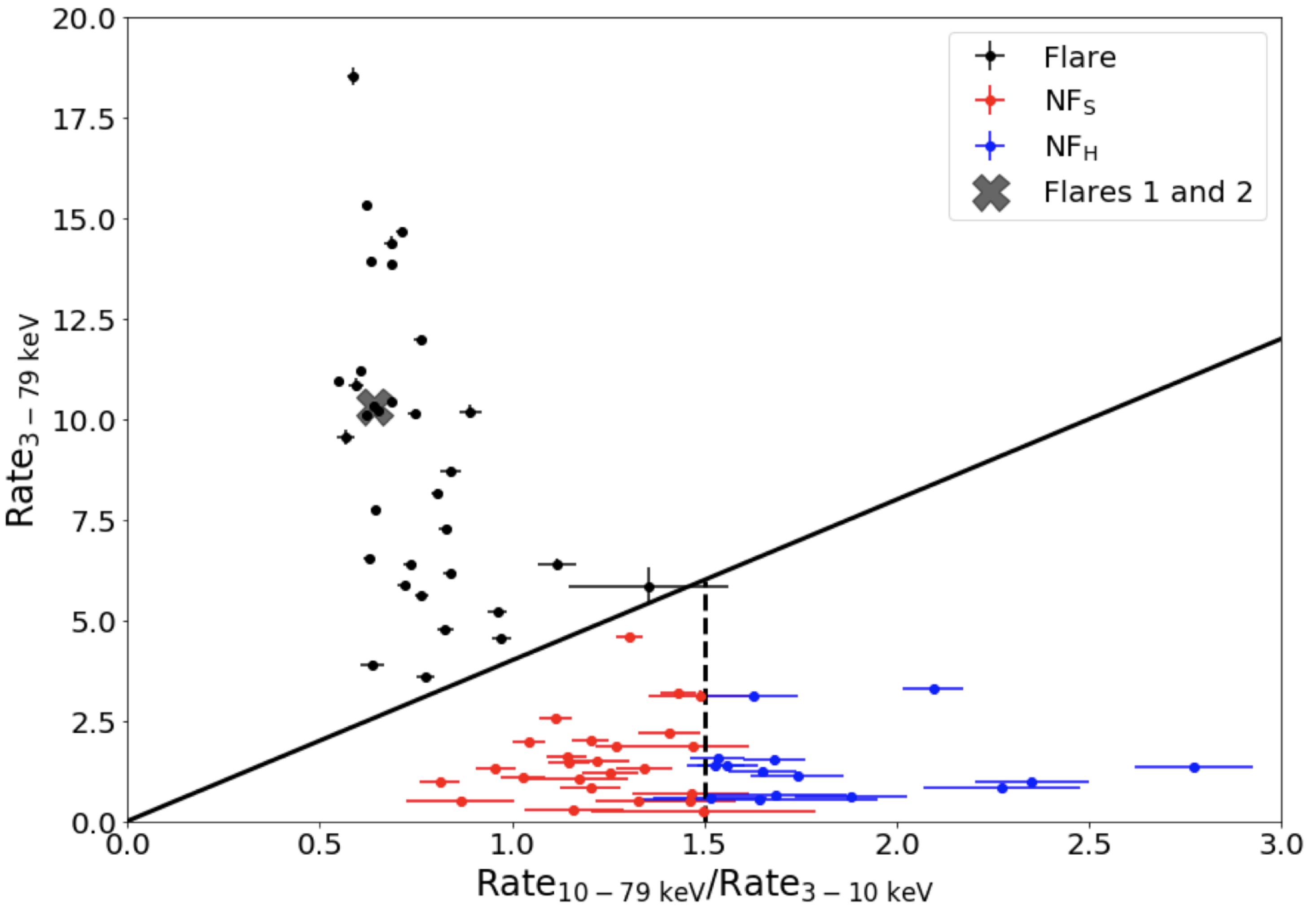}
\caption{The {\sl NuSTAR} hardness-intensity diagram (HID) with 1 ks time bins averaged over the FPMA and B detectors. The solid black line shows the distinction between flaring (black points above the line) and non-flaring (red and blue points below the line) time intervals. The dashed vertical black line shows the further distinction between non-flaring soft (NF$_{\rm S}$; red points to the left of the line) and non-flaring hard (NF$_{\rm H}$; blue points to the right of the line) time intervals (see Sections \ref{v_and_tim} and \ref{rel_ref}). The gray cross shows where the interval containing the two flares simultaneously observed with {\sl NuSTAR} and {\sl NICER} and discussed in Section \ref{nu_nice_flare} lands on the HID.
\label{HID}
}
\end{figure}

\begin{figure}
\centering
\includegraphics[trim={0 0 0 0},width=8.5cm]{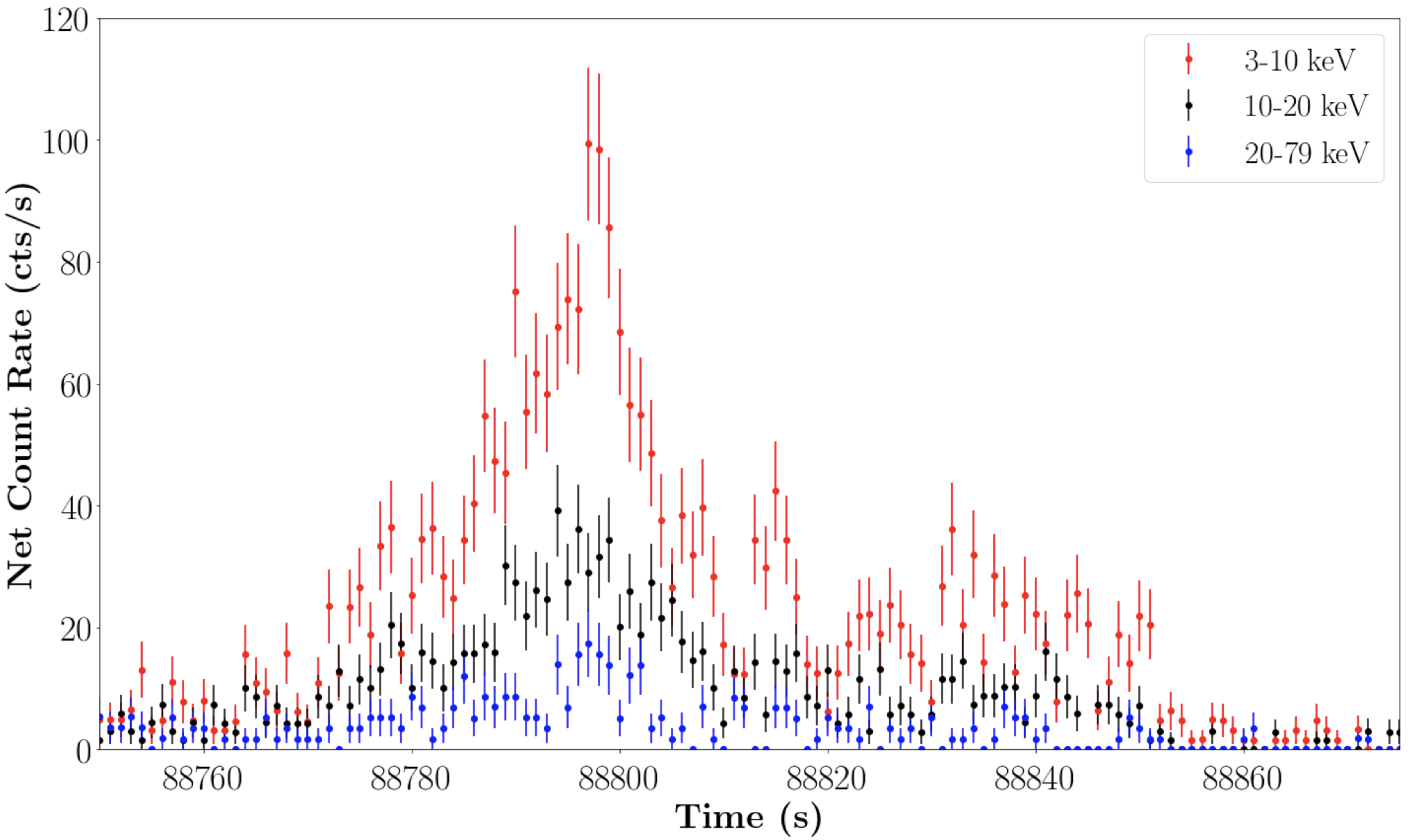}
\caption{The {\sl NuSTAR} FPMA energy resolved (3-10 keV red; 10-20 keV black; 20-79 keV blue) light curves of flare 2 (see Sections \ref{nice_sec} and \ref{nu_nice_flare}) with 1 s time bins. The flares are detected across all energy bands, but have the largest amplitudes at soft (i.e., 3-10 keV) energies.
\label{energy_flare}
}
\end{figure}

To study the variability of Swift J1858 on different time scales, we have produced light curves using a number of different time bins (i.e., 1 s, 10 s, 100 s, 1 ks, 5 ks), which have also been corrected for {\sl NuSTAR}'s various detector effects (e.g., dead time, PSF, vignetting). These light curves reveal that the source was highly variable throughout the {\sl NuSTAR} observation, showing large amplitude flares (see Figure \ref{lc_100s}). These flares typically lasted between $\sim$10-100 s with the largest having a peak count rate $\sim50$ times higher than the source's average count rate (see the inset in Figure \ref{lc_100s}). Throughout the observation, the source also showed changes in its hardness ratio, defined here as the 10-79 keV count rate divided by the 3-10 keV count rate (i.e., $R_{10-79 \rm keV}/R_{3-10 \rm keV}$). The FPMA and B averaged hardness-intensity diagram (HID) shows that the source is softer during the flaring episodes, while spanning a range of hardness ratios during the non-flaring periods (see Figure \ref{HID}). The flares are also observed across {\sl NuSTAR}'s entire band pass, but are most strongly observed in the 3-10 keV energy band (see Figure \ref{energy_flare}).

To better understand the timing properties of the source, and to look for differences between the flaring and non-flaring periods, we use the HID to define two distinct modes of the source, which we designate as either ``flaring" or ``non-flaring". The source is considered to be in the flaring mode (black points in Figure \ref{HID}) when the 3-79 keV energy band count rate and hardness ratio, averaged over 1 ks time intervals, are above the line defined as $R_{3-79 \rm keV}=4\times R_{10-79 \rm keV}/R_{3-10 \rm keV}$ (the black solid line shown in Figure \ref{HID}), while the non-flaring mode includes all of the points below this line (i.e., red and blue points shown in Figure \ref{HID}). The flaring and non-flaring light curves, after being split, consisted of exposure times of $\sim25$ ks and $\sim26$ ks, respectively.

To characterize the observed variability and to search for possible quasi-periodic oscillations (QPOs), we constructed power density spectra\footnote{This observation was not significantly affected by dead time, having a dead time fraction of  $<10\%$. Therefore, we used the typical PDS and not the cross-spectrum (see e.g., \citealt{2015ApJ...800..109B}).} (PDS) from the {\sl NuSTAR} event lists\footnote{We note that Stingray constructs light curves using the event lists, and therefore, does not correct the light curves for the various detector effects mentioned above.} using the Stingray python package \citep{2019arXiv190107681H}. First, light curves with a 4 ms binning were produced from the barycenter corrected {\sl NuSTAR} event files in the 3-79 keV energy range. We also removed 100 s from the beginning and 200 s from the end of each good time interval (GTI) to eliminate any possible effects from an increased background that may occur near the borders of GTIs (see e.g., Section 5 in \citealt{2015ApJ...800..109B}). 

\begin{figure*}
\centering
\includegraphics[trim={0 0 0 0},width=18.0cm]{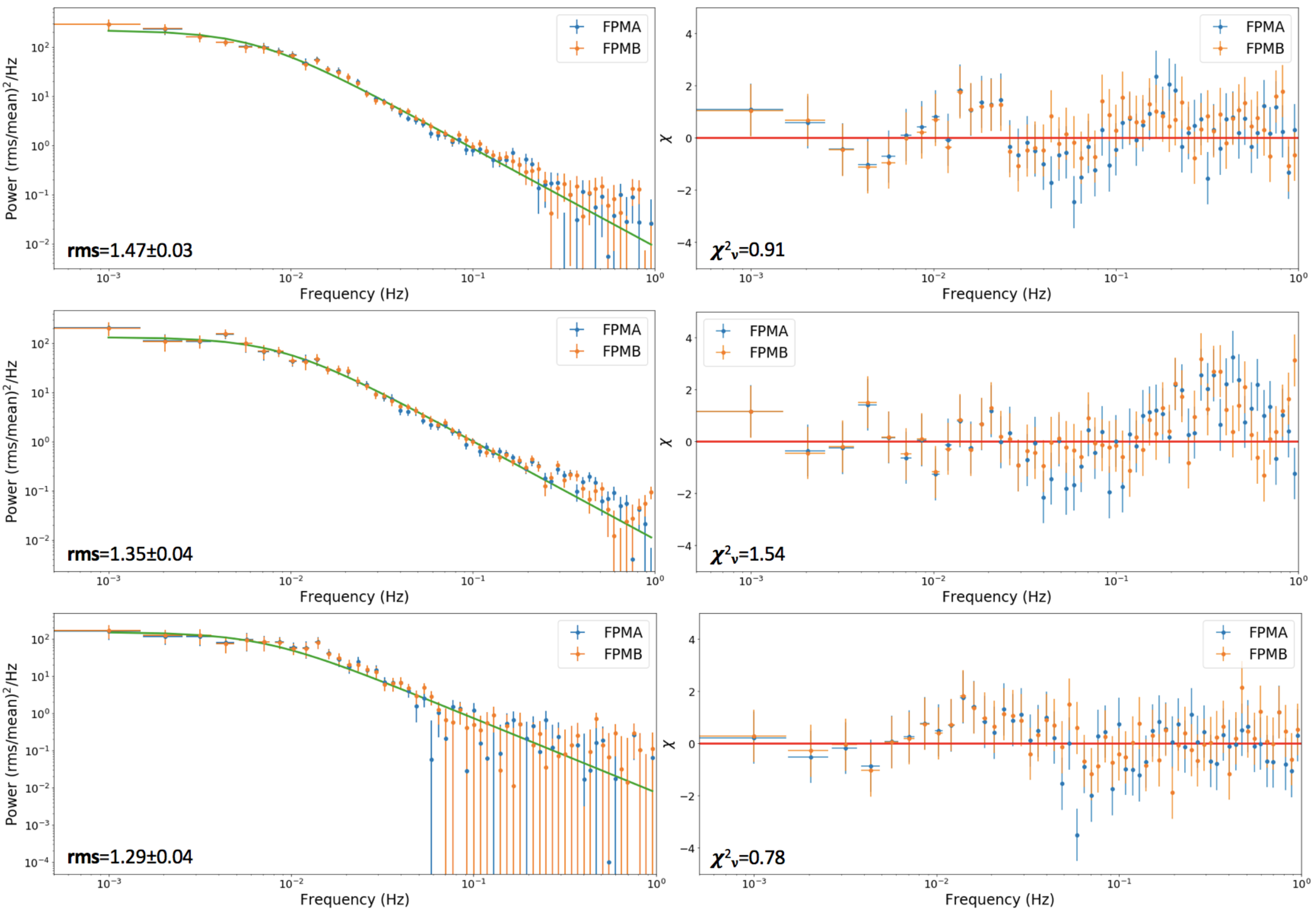}
\caption{{\sl Left:} {\sl NuSTAR} power-spectra of Swift J1858 for the full, flaring, and non-flaring time intervals. The uncertainties in the rms are reported at the 1$\sigma$ level. {\sl Right:} $\Delta\chi$ residuals for the best-fit single, zero frequency centered Lorentzian models.
\label{pow_spec}
}
\end{figure*}

The PDS were produced for the full, flaring, and non-flaring time intervals spanning a 0.001-125 Hz frequency range and were averaged over 1 ks time segments. The PDS were geometrically rebinned by a factor of 1.08 (see Figure \ref{pow_spec}). The error bars in the PDS become very large at frequencies $\gtrsim1$ Hz, so these points were excluded from our analysis after verifying no significant peak is detected in the PDS at these high frequencies. We then simultaneously fit a single, zero frequency centered Lorentzian model to the FPMA and FPMB PDS. The single Lorentzian model fits the data reasonably well, with the exception of the flaring data (see Figure \ref{pow_spec}, right column), and we do not find evidence of any remarkable features (e.g., QPO, orbital periodicity) in the PDS. The best-fit widths of the Lorentzian models, and their 1$\sigma$ uncertainties, are $(1.27\pm0.06)\times10^{-2}$ Hz, $(1.8\pm0.1)\times10^{-2}$ Hz, and $(1.4\pm0.1)\times10^{-2}$ Hz, for the full, flaring and non-flaring time intervals. We derived the fractional rms amplitudes and their $1\sigma$ uncertainties by integrating the best-fit Lorentzian models, which give 147\%$\pm3\%$, 135\%$\pm4\%$, 129\%$\pm4\%$ for the full, flaring, and non-flaring time intervals, respectively. These large rms fractional values are indicative of the large flux fluctuations exhibited by the source.

 The single Lorentzian model is a relatively  poor fit to the flaring PDS ($\chi^2_{\rm red}=1.54$), so we also fit   a model including a second zero frequency centered Lorentzian. This model provides a substantially better fit ($\chi^2_{\rm red}=0.98$), reducing the chi-squared from $\chi^{2}=172.3$ to $\chi^{2}=107.3$ (or $\Delta\chi^{2}=65$) for two fewer degrees of freedom. The best-fit widths for this two Lorentzian model and their 1$\sigma$ uncertainties are (1.51$^{+0.10}_{-0.09})\times10^{-2}$ Hz and ($69^{+12}_{-16})\times10^{-2}$ Hz, for the low (dominating  between 10$^{-3}-10^{-1}$ Hz)  and high (dominating  between 10$^{-1}-1$ Hz) frequency Lorentzians, respectively. The fractional rms amplitudes and their 1$\sigma$ uncertainties are derived in the same way as described in the previous paragraph and are 135\%$\pm3\%$ and 27\%$\pm2\%$ for the low and high frequency Lorentzians, respectively.

\subsection{X-ray Spectra}

\subsubsection{Relativistic Reflection}
\label{rel_ref}

To characterize the spectral differences between J1858's flaring and non-flaring intervals, we extract the energy spectra from three different modes. The first mode is the same as the ``flaring'' mode defined above in Section \ref{v_and_tim} and encompasses the black data points above the solid black line shown in Figure \ref{HID}. Since the non-flaring mode, defined as points below the solid black line in Figure \ref{HID}, spans a broad range of hardness ratios, we further divide this mode by the  dashed black vertical line shown in Figure \ref{HID} into non-flaring soft ($R_{10-79 \rm keV}/R_{3-10 \rm keV}<$1.5) and non-flaring hard ($R_{10-79 \rm keV}/R_{3-10 \rm keV}>$1.5) modes (i.e., shown as red and blue points in Figure \ref{HID}, respectively). The HID bin size of 1 ks was chosen to ensure that the hardness ratio error bars were small enough to confidently differentiate the non-flaring data points into the hard and soft modes\footnote{While it is difficult to accurately separate  the non-flaring soft and hard modes with 100 s binning, it  is still possible to separate the flaring mode data. We have carried  out this exercise and found that the best-fit model is consistent with the  best-fit flaring model shown in Table \ref{tab1} (within uncertainties), with the exception of slightly higher normalizations.}. After dividing the data in this way, exposures of $\sim 25$ ks, $\sim 11$ ks, and $\sim15$ ks remained for the flaring, non-flaring hard, and non-flaring soft modes, respectively. For the remainder of the paper, we denote these three modes as flare, NF$_{\rm S}$, and  NF$_{\rm H}$ for the flaring, non-flaring soft, and non-flaring hard modes, respectively.

\begin{figure*}
\centering
\includegraphics[trim={0 0 0 0},width=18.0cm]{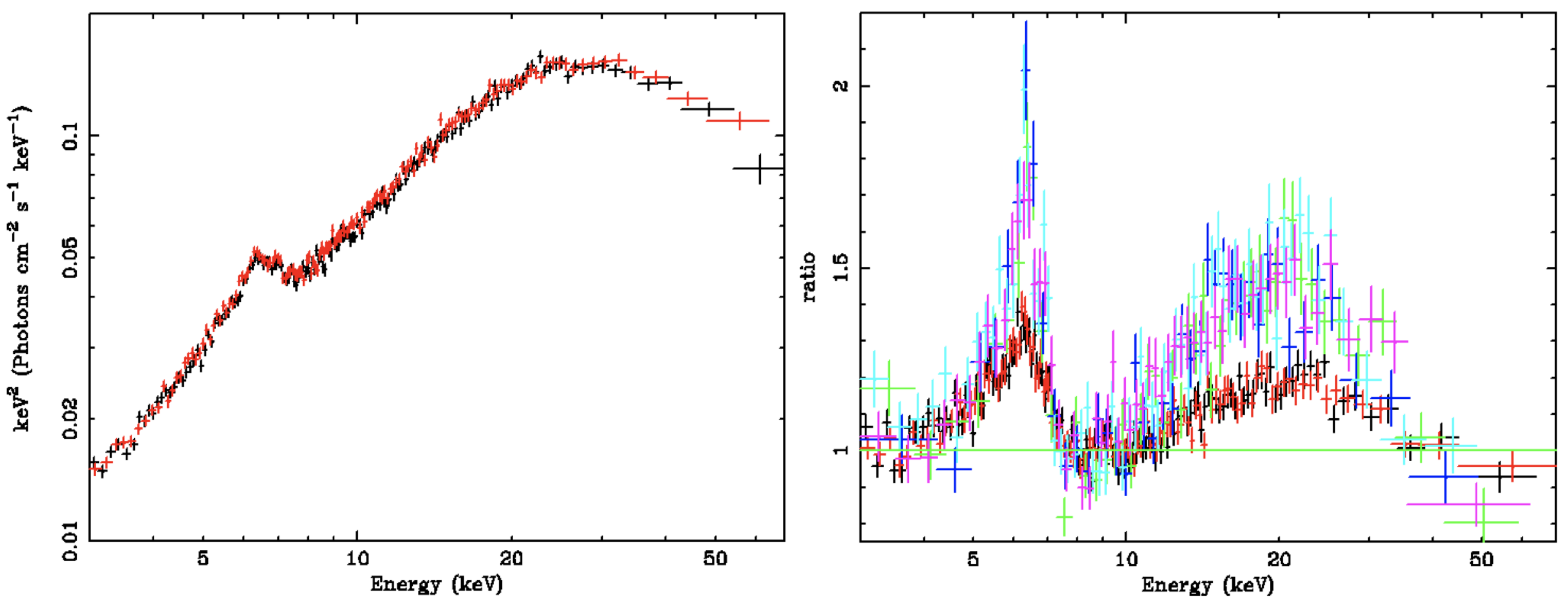}
\caption{ {\sl Left:} Unfolded {\sl NuSTAR} energy spectra of J1858 from the full duration of the observation. {\sl Right:} Data to model ratio for the flaring (red and black), non-flaring hard (green and dark blue) and non-flaring soft (cyan and purple) spectra of J1858 fitted with an exponentially cutoff power-law in the 3-4, 8-10, 30-79  keV energy range and then plotted after reintroducing the data in the full 3-79 keV energy range. All spectra have been rebinned for easier visualization.
\label{raw_spec}
}
\end{figure*}

The {\sl NuSTAR} energy spectra in the 3-79 keV range spanning the entire observation (i.e., not split into the three modes) are shown in Figure \ref{raw_spec} and exhibit a number of features typical of accreting BH systems. These features include an excess of emission at energies between 5-7 keV, typical of iron K features, an absorption edge around 7 keV, and a broad Compton hump above 10-15 keV. To highlight the differences between the X-ray spectra during the three intervals, we fit them with an exponentially cutoff power-law model in the 3-4, 8-10, and 30-79 keV energy bands (i.e., excluding the 4-8 and 10-30 keV energy ranges). These energy bands are chosen as they provide relatively unbiased access to the underlying continuum by avoiding the strongest spectral features previously mentioned (see e.g., \citealt{2017ApJ...839..110W}). We then plot the ratio of the data to the folded model,  after re-including the data in the full 3-79 keV energy range, in the right panel of Figure \ref{raw_spec}.  Clear differences can be seen between the spectra from the three time intervals. For instance, the iron lines observed in the NF$_{\rm S}$ and NF$_{\rm H}$ spectra are more strongly peaked around 6.4 keV, while also having a more pronounced absorption edge at $\sim$7 keV  and Compton hump above $\sim10$ keV when compared to the flaring spectrum. All three spectra show evidence of a red wing, i.e., a broadening of the iron line extending to lower energies. The red wing provides strong evidence that this emission is coming from relativistically broadened reflection of photons off the innermost regions of the accretion disk (see e.g., \citealt{1989MNRAS.238..729F,1991ApJ...376...90L}). Lastly, the Compton hump emission is most pronounced in the NF$_{\rm H}$ spectra.

Given the broad reflection features evident in the X-ray spectra, we use the collection of RELXILL models (version 1.2.0; \citealt{2014ApJ...782...76G}) to fit them. The RELXILL models combine the XILLVER  \citep{2010ApJ...718..695G} reflection model with the RELCONV relativistic convolution model, which captures the relativistic effects due to the emitting material's close proximity to the BH \citep{2010MNRAS.409.1534D}. In particular, we use the RELXILLLPCP model to fit the X-ray spectra of J1858. This model uses the thermally Comptonized continuum model {\tt nthComp} \citep{1996MNRAS.283..193Z,1999MNRAS.309..561Z} for the input continuum spectrum and assumes a lamp-post geometry (i.e., a point source directly above the spin axis of the BH) for the illuminating X-ray source.

RELXILLLPCP is characterized by several physical parameters intrinsic to the BH binary. This includes the inclination of the inner accretion disk, $i$, the iron abundance of the accreted material, $A_{\rm Fe}$, and the spin of the BH, $a^{*}$. Since these parameters are not expected to change during the duration of our observation, they are linked between all spectra during the fitting procedure. This model also contains parameters which can change between the different time intervals, including the photon-index of the power-law emission incident on the accretion disk, $\Gamma$, the temperature of the electrons in the corona, $kT_{\rm e}$, the height of the source above the BH that is irradiating the accretion disk, $h$, and the ionization state of the iron in the accretion disk, $\log{\xi}$\footnote{Here $\xi=4\pi F_X/n$, where $F_X$ is the ionizing flux incident on the accretion disk, and $n$ is the density of the material in the disk.}.  These parameters are allowed to vary for the flaring, NF$_{\rm S}$, and NF$_{\rm H}$ spectra during our fits. The reflection fraction, $R_{\rm refl}$, is another parameter in this model and is defined as the ratio of the amount of light from the primary source that is emitted towards the disk versus the amount that escapes to infinity \citep{2016A&A...590A..76D}. For the lamp-post geometry, the RELXILL package offers the option to calculate the reflection fraction in a self-consistent way using relativistic ray-tracing, which we take advantage of for our fits. Lastly, the model contains the inner radius of the accretion disk, $R_{\rm in}$, which we allow to vary across all three spectral modes.

\begin{figure*}
\centering
\includegraphics[trim={20 0 0 0},width=17cm]{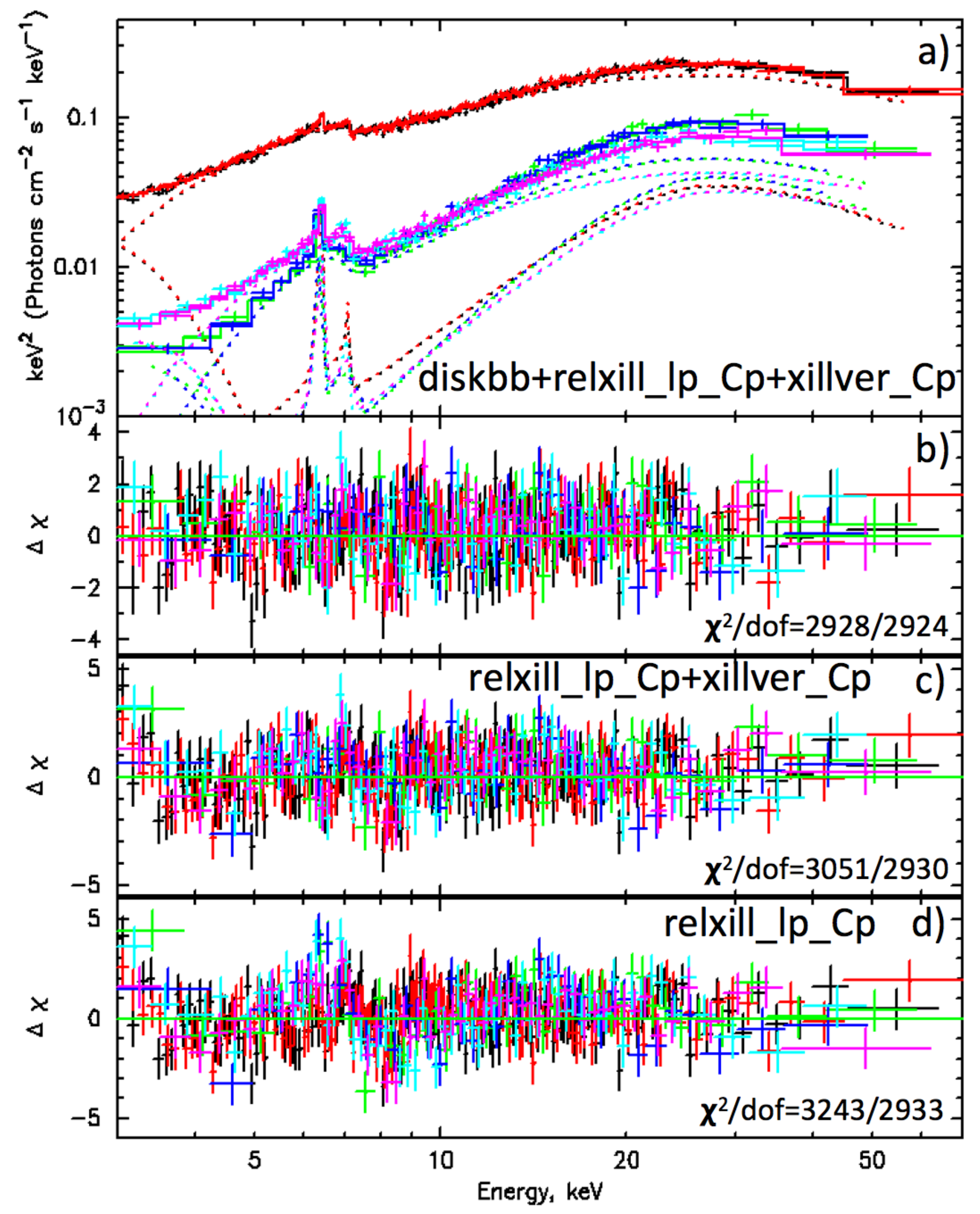}
\caption{ a) {\sl NuSTAR} energy spectra of J1858 from the flaring (red and black) non-flaring hard (green and blue), and non-flaring soft (cyan and magenta) time intervals. The best fit model components from the {\tt relxill\_lp\_cp$+$xillver\_cp$+$diskbb} model are plotted as dashed-lines in each corresponding color.  b)  Residuals of the data to the model for the best-fit model.  c)  Residuals of the data to the model for the {\tt relxill\_lp\_cp$+$xillver\_cp} model. Note that large residuals appear at soft X-ray energies.  d) Residuals of the data to the model for the {\tt relxill\_lp\_cp} model.  Large residuals appear at soft X-ray energies and around the iron line complex. All spectra have been rebinned for easier visualization.
\label{best_spec}
}
\end{figure*}

The best-fit RELXILLLPCP model to the data has a reduced chi-squared of $\chi^2/\nu=3243/2933$ and still shows large residuals both at low energies and near the iron complex around 6.4 keV (see Figure \ref{best_spec}d). The iron line features are more strongly peaked in the residuals of the NF$_{\rm S}$ and NF$_{\rm H}$ spectra, and are possibly due to reflection from distant cold material. This suggests that additional model components are needed to adequately fit the spectra. To account for the excess near the iron complex, we add a neutral (i.e., $\log{\xi}=0$)  XILLVERCP component to the model. The XILLVERCP model was chosen because its continuum emission model (i.e., {\tt nthComp}) is the same as that used in the RELXILLLPCP model. Therefore, we tie the parameters shared by both models together as the same source should be illuminating both the accretion disk and distant reflector\footnote{Realistically, the distant reflector will see the illuminating source  gravitationally redshifted. However, due to the relatively large illuminating source heights, the gravitational redshift is small ($\lesssim0.25$),  making this  effect negligible.}.  Additionally, we fix the ionization of the distant reflector to $\log\xi=0$ because the narrow part of the residual is peaked near the 6.4 keV Fe K$\alpha$ line, implying that it is likely coming from neutral iron. We also assume that all of the emission coming from this component of the model is reflected (i.e., denoted in the XILLVERCP model by setting the reflection fraction to -1). We allow the normalization of the XILLVERCP component to vary between the flaring,  NF$_{\rm H}$, and NF$_{\rm S}$ spectra. Following this addition, the reduced chi-squared, $\chi^2/\nu=3051/2930$, improved by $>190$ for three fewer degrees of freedom. However, the excess at soft X-ray energies still remains (see Figure \ref{best_spec}c), so we add a multi-temperature blackbody ({\tt diskbb}; \citealt{1984PASJ...36..741M}) to account for thermal emission from the accretion disk, which is likely causing this excess. We allow the accretion disk temperature and normalization of the DISKBB component to vary between the flaring,  NF$_{\rm H}$, and NF$_{\rm S}$ spectra. Hence, the complete model is {\tt const*tbnew*(diskbb$+$relxill\_lp\_cp$+$xillver\_cp)}. This additional component further improved the fit, leading to an improvement of $\sim120$ in the reduced chi-squared ($\chi^2/\nu=2928/2924$) for six fewer degrees of freedom. It is interesting to note that about half of the reduction in the chi-squared comes from the addition of the thermal component to the flaring spectrum.  The best-fit model parameters for J1858 can be found in Table \ref{tab1}, while the best-fit spectra and their residuals are shown in Figures \ref{best_spec}a and \ref{best_spec}b, respectively.

Using the best fit model, we have also tried to require the absorbing column density to be the same across the spectra of all three source modes. This led to a best-fit absorbing column density of $N_{\rm H}=18\times10^{22}$ cm$^{-2}$ and a reduced chi-squared of $\chi^2/\nu=2958/2926$. This fit has an additional $\Delta\chi^{2}=30$ for two additional degrees of freedom. We disfavor this model because Figure \ref{raw_spec} shows clear differences in the strength of the absorption edge at $\sim7$ keV, strongly suggesting that there is additional intrinsic  absorption during the non-flaring mode compared to  the flaring mode (see e.g., \citealt{2017ApJ...839..110W}).

\begin{table*}
\caption{Best-fit parameters of the {\tt const*tbnew(diskbb$+$relxill\_lp\_cp$+$xillver\_cp)} model simultaneously fit to the flaring, non-flaring hard (NF$_{\rm H}$), and non-flaring soft (NF$_{\rm S}$) {\sl NuSTAR} spectra of Swift J1858.} 
\label{tab1}
\begin{center}
\renewcommand{\tabcolsep}{0.11cm}
\begin{tabular}{lccccc}
Model component & Parameter & Units & Flare & NF$_{\rm H}$ & NF$_{\rm S}$\\
\tableline
Constant & ... & FPMB/FPMA & 1.020$\pm0.007$ & 1.02$\pm0.02$ & 1.02$\pm0.02$ \\
\tableline
TBNEW & $N_{\rm H}$ & $10^{22}$ cm$^{-2}$ &   14$\pm2$ & 42$^{+8}_{-9}$ & 30$^{+8}_{-7}$\\
\tableline
DISKBB & $T_{\rm in}$ & keV &  0.36$^{+0.03}_{-0.05}$ & 0.31$\pm0.03$ & 0.35$\pm0.04$ \\
 & Norm & 10$^{3}\times$($R_{\rm in,km}\tablenotemark{d}/D_{10}\tablenotemark{e})^{2}\cos{i}$ & 7$^{+20}_{-4}$ & 62$^{+97}_{-51}$ & 8$^{+13}_{-6}$\\
 \tableline
 RELXILLLPCP & $i\tablenotemark{b}$ & degrees & $<23^{\circ}$ & -- & -- \\
   & $A_{\rm Fe}\tablenotemark{b}$ & solar & 1.0$^{+0.4}_{-0.2}$ & -- & -- \\
   & $\Gamma$ & ... & 1.50$\pm0.03$ & 1.41$^{+0.10}_{-0.08}$ & 1.49$^{+0.1}_{-0.03}$ \\
   & $kT_{e}$ & keV & 15.0$^{+0.9}_{-0.7}$ & 14$\pm2$ & 16$\pm2$ \\
   & $h$ & $r_g$ & 13$^{+7}_{-5}$ & 6$^{+6}_{-4}$ & 5$^{+3}_{-2}$ \\
   & $a^{*}\tablenotemark{b}$ & ... & $>0.0$ & -- & -- \\
   & $\log{\xi}$ & log (erg cm s$^{-1}$) & 3.54$^{+0.10}_{-0.13}$ & 3.0$\pm0.3$ & 3.2$^{+0.1}_{-0.2}$ \\
   & $r_{\rm in}$ & $r_{\rm ISCO}$ & $<8$ & $<6$ & $<5$  \\
   & $R_{\rm refl}$ & ... & 1.3\tablenotemark{a} & 1.6\tablenotemark{a} & 1.5\tablenotemark{a}  \\
   & Norm & 10$^{-4}$ & 6.6$^{+1.2}_{-0.7}$ & 2.5$^{+3.3}_{-0.9}$ & 2.6$^{+1.7}_{-0.8}$  \\
   \tableline
   XILLVERLPCP & $\log{\xi}$ & log (erg cm s$^{-1}$) & 0.0\tablenotemark{c} & 0.0\tablenotemark{c} & 0.0\tablenotemark{c} \\
   & $R_{\rm refl}$ & ... & -1\tablenotemark{c} & -1\tablenotemark{c} & -1\tablenotemark{c} \\
   & Norm & $10^{-4}$ & 2.0$\pm0.7$ & 2.0$\pm0.9$ & 1.8$^{+0.7}_{-0.5}$ \\
  \tableline
  Observed Flux & 3.0-79 keV & $10^{-10}$ erg cm$^{-2}$ s$^{-1}$ & 6.79$\pm0.02$ & 2.10$\pm0.02$ & 1.88$\pm0.02$ \\
  Unabsorbed Flux & 3.0-79 keV & $10^{-10}$ erg cm$^{-2}$ s$^{-1}$ & 7.40$\pm0.02$ & 2.56$\pm0.03$ & 2.19$\pm0.02$ \\
    \tableline
   $\chi^{2}$/d.o.f. & 2928/2924 & 1.001 \\
   \tableline
\end{tabular} 
\tablenotetext{a}{Calculated self-consistently by the RELXILLLPCP model.}
\tablenotetext{b}{Parameter is tied across Flare, NF$_{\rm S}$, and NF$_{\rm H}$ spectral models. The best-fit value is given in the Flare column.}
\tablenotetext{c}{Fixed value.}
\tablenotetext{d}{The apparent inner disk radius in units of km.}
\tablenotetext{e}{Distance to the source in units of 10 kpc.}
\end{center}
\end{table*}

\subsubsection{Joint NICER and NuSTAR fits}
\label{nu_nice_flare}
The relativistic reflection fits to the {\sl NuSTAR} energy spectra place constraints on several spectral features that are more prominently observed in the soft X-ray band (e.g., large intrinsic absorbing column density, {\tt diskbb} temperature and normalization).  Therefore, to verify that the values of these features derived from the {\sl NuSTAR} spectra are reasonable, we use the simultaneous {\sl NICER} observations\footnote{The {\sl Neil Gehrels Swift Observatory}'s X-ray Telescope (XRT) also observed J1858 simultaneously with {\sl NuSTAR}  for $\sim400$ s (obsID 00010955002). However, due to the short duration of the observation, no bright flares from J1858 were detected by {\sl Swift}-XRT leaving only $\sim250$ total counts.}. To minimize the effects of {\sl NICER}'s background contribution to the X-ray energy spectra we chose to focus on short ($\sim100$ s), flares simultaneously observed by {\sl NuSTAR} and {\sl NICER}.

The energy spectra from the first flare, which reached peak count rates of $\sim770$ cts s$^{-1}$ and $\sim85$ cts s$^{-1}$ for {\sl NICER}  and {\sl NuSTAR}, respectively, were extracted from a 60 s time interval around the flare. These spectra contain a total of $\sim950$ counts in each {\sl NuSTAR} focal plane module and $\sim11,100$ counts in {\sl NICER}. The energy spectra of the second flare, reaching peak {\sl NuSTAR} and {\sl NICER} count rates of $\sim135$ cts s$^{-1}$ and $\sim860$ cts s$^{-1}$, were extracted from a 100 s window containing the flare (see the inset in Figure \ref{lc_100s}). This flare's spectrum contained $\sim2,300$ counts in each {\sl NuSTAR} focal plane module and $\sim17,000$ counts in {\sl NICER}.  These flares have a 1 ks average {\sl NuSTAR} count rate of $\sim 10$ cts s$^{-1}$ and a hardness ratio of $\sim 0.6$ (see the black cross in Figure \ref{HID}). Prior to fitting the spectra we grouped them to have 100 counts per energy bin for the {\sl NICER} spectra and 50 counts per energy bin for the {\sl NuSTAR} spectra. Unfortunately, there appears to be strong systematic residuals in the {\sl NICER} spectra around 0.5 keV, so we avoid energies below 0.6 keV in our fits.

\begin{figure}
\centering
\includegraphics[trim={0 0 0 0},width=9.0cm]{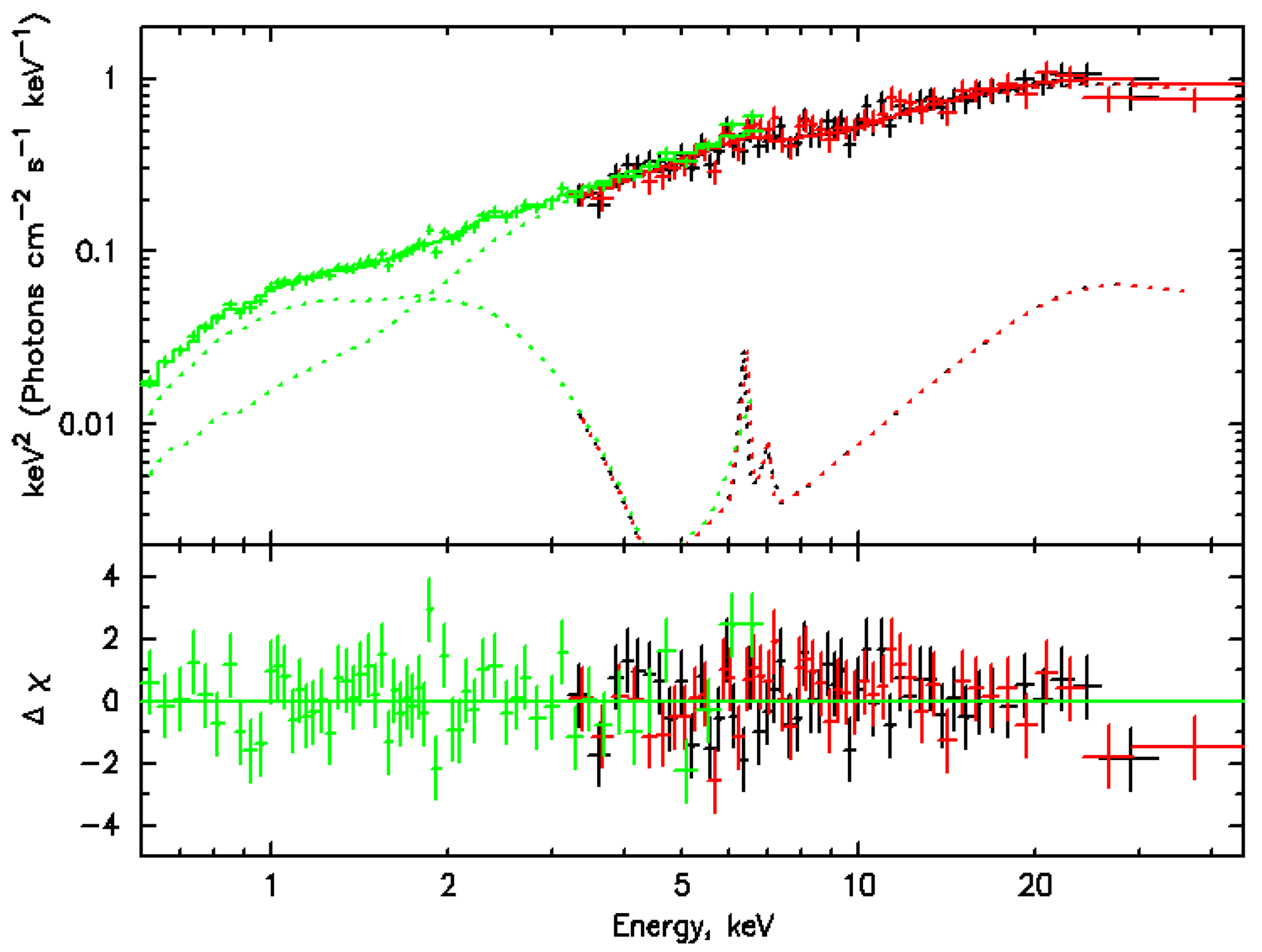}
\caption{ The best-fit {\tt const*tbnew*pcfabs(diskbb$+$relxill\_lp\_cp$+$xillver\_cp)} {\sl NICER} (green) and {\sl NuSTAR} (black and red for FPMA and FPMB, respectively) energy spectra. The {\sl NICER} data were limited to the 0.6-7 keV band (see Section \ref{nu_nice_flare}). The best-fit model parameters are  shown in Table \ref{tab2}.
\label{nu_nice_spec}
}
\end{figure}

To check that the large intrinsic absorption and thermal component observed by {\sl NuSTAR} is consistent with the {\sl NICER} spectra, we jointly fit the {\sl NICER} and {\sl NuSTAR} spectra with the best-fit flaring model shown in Table \ref{tab1}.  Due to the relatively small number of counts in the {\sl NuSTAR} energy spectra for these short duration flares we freeze  all of the parameters of the model except for the absorbing column density, the normalizations for the three model components (i.e., {\tt rellxill\_lp\_cp}, {\tt xillver\_lp\_cp}, {\tt diskbb}), and the cross-calibration constants\footnote{For simplicity, the inner-radius of the accretion disk is frozen at the ISCO (i.e., $R_{\rm in}=R_{\rm ISCO}$) for these fits.}. This model produces a very poor fit to the spectra of flares 1 and 2, having reduced chi-squared values of $\chi^2/\nu=200/133$ and $\chi^2/\nu=675/224$, respectively, and giving small absorbing column densities of $N_{\rm H}\approx4\times10^{21}$ cm$^{-2}$. However, the fits substantially improve if a partially covering absorber ({\tt pcfabs}) is added to the model and allowed to vary ($\chi^2/\nu=129/131$, and $\chi^2/\nu=209/222$, for flares 1 and 2, respectively). The best-fit parameters of the model to the two flares are shown in Table \ref{tab2}, while the best-fit spectra  and residuals  for flare 2  are shown in Figure \ref{nu_nice_spec}. Most notably, the {\tt diskbb} normalization is consistent with the {\sl NuSTAR} only fits for both flares 1 and 2, suggesting that the thermal component required by the fits to the {\sl NuSTAR} only data is confirmed by the {\sl NICER} data.

\begin{table*}
\caption{Best-fit parameters of the {\tt const*tbnew*pcfabs(diskbb$+$relxill\_lp\_cp$+$xillver\_cp)} model fit to flares 1 and 2, which were simultaneously observed by {\sl NuSTAR} and {\sl NICER}.} 
\label{tab2}
\begin{center}
\renewcommand{\tabcolsep}{0.11cm}
\begin{tabular}{lcccc}
Model component & Parameter & Units & Flare 1& Flare 2\\
\tableline
Constant & ... & FPMB/FPMA & 1.12$\pm0.08$ & 1.00$\pm0.05$ \\
Constant & ... & {\sl NICER}/FPMA & 1.18$\pm0.08$ & 1.07$\pm0.05$ \\
\tableline
TBNEW & $N_{\rm H}$ & $10^{22}$ cm$^{-2}$ &   0.41$\pm0.03$ & 0.39$\pm0.02$ \\
\tableline
PCFABS & $N_{\rm H}$ & $10^{22}$ cm$^{-2}$ &   5.3$^{+1.1}_{-0.9}$ & 6.2$^{+0.7}_{-0.5}$ \\
& Covering Fraction &  &   0.78$^{+0.09}_{-0.21}$ &  0.84$^{+0.05}_{-0.09}$ \\
\tableline
DISKBB & Norm & 10$^{3}\times$($R_{\rm in,km}/D_{10})^{2}\cos{i}$ & 4$^{+4}_{-3}$ & 5$^{+3}_{-2}$ \\
 \tableline
 RELXILLLPCP & Norm & 10$^{-4}$ & 24$\pm1$ & 32$\pm2$  \\
  \tableline
   XILLVERLPCP & Norm & $10^{-4}$ & $<2.5$ & $<9$ \\
  \tableline
  Observed Flux & 0.6-79 keV & $10^{-9}$ erg cm$^{-2}$ s$^{-1}$ & 2.52$\pm0.04$ & 3.46$\pm0.04$  \\
 Unabsorbed Flux & 0.6-79 keV & $10^{-9}$ erg cm$^{-2}$ s$^{-1}$ & 3.51$\pm0.05$ & 4.78$\pm0.05$  \\
    \tableline
  $\chi^{2}$/d.o.f. & & & 129/130 & 209/222 \\
   \tableline
\end{tabular} 
\tablenotetext{}{Model parameters not listed here were fixed to the best fit values shown in Table \ref{tab1}.}
\end{center}
\end{table*}

\section{Discussion}
\label{diss}

\subsection{Spectral and Timing Features}
\begin{figure}
\centering
\includegraphics[trim={0 0 0 0},width=9.0cm]{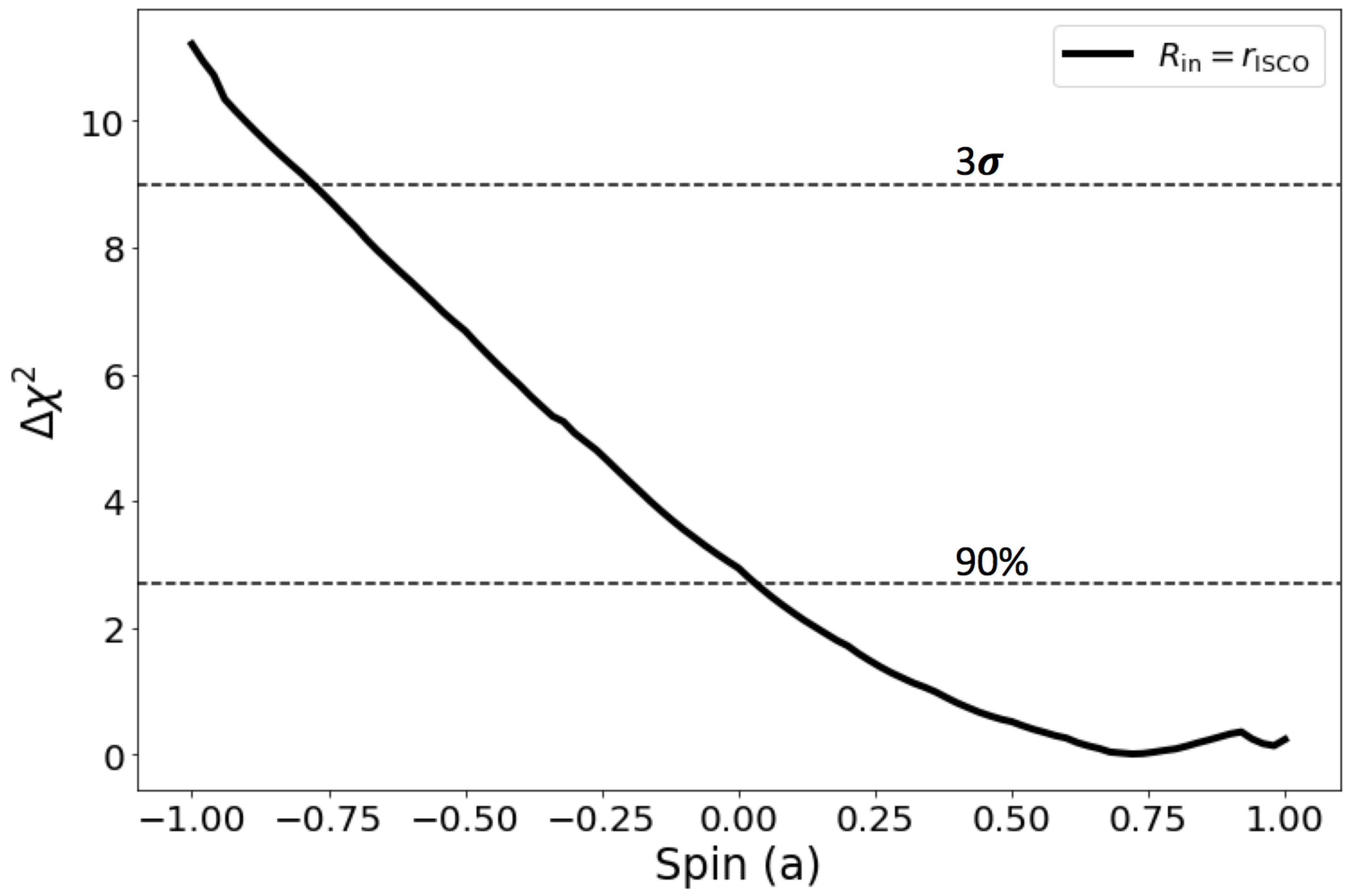}
\caption{ Constraints on the lower-limit of the spin parameter of J1858 using Xspec's {\tt steppar} command. The dashed black lines show the 90\% and 3$\sigma$ confidence intervals for a single parameter. The solid black line shows the $\Delta \chi^{2}$ of the spin parameter if the inner-radius of the accretion disk is frozen at the ISCO.
\label{spin_stepp}
}
\end{figure}

The best-fit model parameters of the physical values inherent to the binary are an iron abundance of the accreting material, $A_{\rm Fe}=1.00^{+0.4}_{-0.2}$, and a 90\% upper-limit on the inclination of the inner accretion disk $i<23^{\circ}$.  We also calculate the $3\sigma$ upper-limit on the inclination, finding $i<29^{\circ}$. This suggests that the disk is viewed almost face on. We also find 90\% upper-limits for the inner-radius of the accretion disk, $R_{\rm in}$, of $<8$ $r_{\rm ISCO}$, $<6$ $r_{\rm ISCO}$, and $<5$ $r_{\rm ISCO}$ for the flaring, NF$_{\rm H}$, and NF$_{\rm S}$ spectra, respectively, which are consistent with  $R_{\rm in}=R_{\rm ISCO}$. It is important to note that the spin and $R_{\rm in}$ are degenerate with each other. Therefore, to calculate a lower-limit on the spin of the BH, we set $R_{\rm in}$ to be at $r_{\rm ISCO}$, refit the model, and calculate the lower-limit on the spin. After fitting, we find that all best-fit parameters in this updated model are consistent (within the 90\% uncertainties) with those found previously and shown in Table \ref{tab1}. This model gives  a 90\% confidence lower-limit on the spin of $a^{*}>0.0$ and a 3$\sigma$ lower-limit of $a^{*}>-0.8$ (see Figure \ref{spin_stepp}).

One of the most striking features of the best-fit model to the {\sl NuSTAR} energy spectra is the absorbing column density, which is very large even in the flaring spectrum (i.e., $N_{\rm H}=14\pm2\times10^{22}$ cm$^{-2}$). This absorption is a factor of $\sim2-3$ larger for the NF$_{\rm H}$ ($N_{\rm H}=42^{+8}_{-9}\times10^{22}$ cm$^{-2}$) and NF$_{\rm S}$ ($N_{\rm H}=30^{+8}_{-7}\times10^{22}$ cm$^{-2}$) spectra. However, this absorption cannot fully account for the change in the source's flux (see unabsorbed fluxes in Table \ref{tab1}), implying that the source must also be intrinsically variable. Furthermore, the absorption appears to be anti-correlated with the source's intrinsic (i.e., unabsorbed) flux between the flaring, NF$_{\rm H}$, and NF$_{\rm S}$ modes. On the other hand, the intrinsic absorption found in our joint {\sl NICER} and {\sl NuSTAR} fits is about a factor of two smaller than the intrinsic absorption found from the fits to the averaged flaring {\sl NuSTAR} spectra. Assuming that the intrinsic absorption is caused by dense clouds of material intersecting the observer's line-of-sight, similar to V404 Cyg (see e.g., \citealt{2017MNRAS.468..981M}; and Section \ref{comp_v404} for more details) this difference is somewhat expected. This is because flaring modes are defined in the {\sl NuSTAR} data using the 1 ks binned light curves. However, given the rapid variability of the source, this binning will inevitably include some short periods when the source is not flaring\footnote{See the inset in Figure \ref{lc_100s} for an example of a typical 1 ks flaring bin viewed on shorter 1 s timescales.}. Therefore, since these non-flaring periods are correlated with a larger intrinsic absorption, then the 1 ks ``averaged" absorbing column density would be expected to be larger than the absorbing column density observed during a shorter $\sim100$ s flare. Furthermore, in a later {\sl Swift}-XRT observation (i.e., not the one coincident with our {\sl NuSTAR} observation) it was found that energy spectrum at soft X-ray energies could also be adequately fit by a similar partially covered thermal plasma plus a power-law model, with the partially covering absorber having a significant absorbing column density of $N_{\rm H}=17^{+0.6}_{-0.5}\times10^{22}$ cm$^{-2}$ \citep{2018ATel12220....1R}, which is consistent with the value found from the fits to the flaring {\sl NuSTAR} energy spectra\footnote{The {\sl Swift} spectra were not divided into flaring and non-flaring modes for this spectral fit.}. Additional support for the fact that most of this absorption must be intrinsic to the binary itself, is that the total Galactic HI absorbing column density in this direction is only $N_{\rm H}\approx1.8\times10^{21}$ cm$^{-2}$ \citep{2016A&A...594A.116H}. We also note that the Galactic HI absorbing column density is consistent (within a factor of about two) with the value found for the interstellar absorption in our joint {\sl NICER}$+${\sl NuSTAR} spectral fits (i.e., $N_{\rm H}\approx4\times10^{21}$ cm$^{-2}$).

Although no QPOs or orbital modulation were found in our X-ray light curves or PDS they did show large amounts of variability. The full PDS had a large fractional rms amplitude of 1.47 (see Section \ref{v_and_tim}), capturing the large flares observed in the light curves. It should be noted that there has been a report of a relatively low frequency QPO in a later {\sl NuSTAR} observation at a frequency of 2.7$\times10^{-3}$ Hz (i.e., $\sim$364 s; \citealt{2019ATel12512....1H}), suggesting that this QPO unfortunately formed while the source was Sun constrained for {\sl NuSTAR}. The analysis of this later {\sl NuSTAR}  observation and potential QPO will be presented in a future publication.

\subsection{Optical Counterpart and Distance}

An optical/UV counterpart coincident with J1858's {\sl Swift}-XRT position was first detected by {\sl Swift}-UVOT, having a UVW2 Vega magnitude of 17.38$\pm0.08$ \citep{2018ATel12160....1K}. Additional follow-up observations found that this optical source was highly variable, varying between $r'$ magnitudes of $\sim18.4-16.3$, with variations as large as 1 magnitude on timescales of two minutes to as short as $<5$ s (\citealt{2018ATel12164....1V,2018ATel12180....1B,2018ATel12186....1R,2018ATel12197....1P}). Further, the slower, but brighter flares were found to be more blue, while the fast flares were found to be more red \citep{2018ATel12197....1P}. Interestingly, the optical counterpart was bright enough to be detected by Pan-STARRs, with an $r'$ magnitude of 19.97$\pm0.03$ prior to the source going into outburst\footnote{We note that the Pan-STARRs observations were obtained in 2012, long before the source was observed to be in outburst \citep{2016arXiv161205560C}.} \citep{2016arXiv161205560C,2018ATel12160....1K}. This suggests that the optical source will still likely be detectable after J1858 returns to quiescence, allowing for follow-up optical/near-infrared spectroscopy to determine the spectral type of the companion star. It may also be possible to constrain the binary orbital period and inclination once the source returns to quiescence. 

The optical counterpart has also shown strong P-Cygni profiles in multiple emission lines, suggesting the system contains a strong optical wind \citep{2019ATel12881....1M}. The terminal velocity of the wind was found to be $\sim2,500$ km/s. The features observed in these emission lines were also found to vary on $\sim5$ minute timescales \citep{2019ATel12881....1M}. There is also evidence that the terminal wind velocity has changed in magnitude over time, with earlier observations having smaller wind velocities of $\sim500-1,500$ km/s \citep{2019ATel12881....1M}.

It is interesting to note that the source was detected in optical prior to its outburst, and in UV during the outburst, suggesting that the source is likely to be relatively nearby. There appears to be a low amount of extinction (i.e., $E(B-V)\approx0.3$ out to a distance of $\sim4$ kpc) in this direction from the \cite{2016ApJ...818..130B} extinction maps, but unfortunately the maps do not extend beyond $\sim4$ kpc. Therefore, it is difficult to set any constraints on the source distance at this time. If, however, we assume a fiducial distance of $\sim5$ kpc, then the unabsorbed X-ray luminosity of flare 2 (see Table \ref{tab2}) reaches $L_X=1.4\times10^{37}(d/{\rm 5 \ kpc})^{2}$ erg s$^{-1}$ in the 0.6-79 keV energy band. We find an unabsorbed peak X-ray luminosity for flare 2 of $L_X\approx7\times10^{37}(d/{\rm 5 \ kpc})^{2}$ erg s$^{-1}$ in the 0.6-79 keV energy band as the peak count rate for flare 2 is about a factor of 5 larger than the average flare count rate. This luminosity is about $5\%$ of the Eddington luminosity assuming a 10 solar mass BH. Therefore, this system would need to be at a distance of $\gtrsim20$ kpc to be accreting near the Eddington limit. Moreover, optically bright BH transients have been found to lie at a median distance of $\sim2$ kpc \citep{2019MNRAS.485.2642G}, suggesting that distances $>20$ kpc are somewhat unlikely.

\subsection{Comparison with similar systems}
\label{comp_v404}

J1858's flaring behavior in the X-ray through optical bands\footnote{We note that rapid X-ray flaring has also been observed in several other BHBs, such as GRS 1915+105 and IGR J1709-3624 (see e.g., \citealt{2016Natur.529...54K}).}, lack of a well defined spectral state, large intrinsic absorbing column density, and winds detected in the optical band are reminiscent of the well known sources V404 Cyg and V4641 Sgr.  V404 Cyg, similar to J1858, showed rapid and large amplitude flares in its X-ray light curves during its outbursts in 1989 and more recently in 2015 (see e.g., \citealt{1989Natur.342..518K,1999MNRAS.309..561Z,2017ApJ...839..110W}). In fact, during V404 Cyg's  2015 outburst, the hardness ratios of a majority of the flares were found to be around $R_{10-79 \rm keV}/R_{3-10 \rm keV}\approx0.5$, similar to what is observed for J1858 (see Figure 3 in \citealt{2017ApJ...839..110W}). Additionally, V404 Cyg showed dramatic changes in the intrinsic absorbing column density, reaching values as large as $N_{\rm H}\approx10^{25}$ cm$^{-2}$, which were anti-correlated with the flux of the source \citep{2017MNRAS.468..981M,2017MNRAS.471.1797M}. At optical wavelengths, slow blue flares and fast red flares, like those observed in J1858, were also exhibited by V404 Cyg \citep{2016MNRAS.459..554G}. The detection of varying P-Cygni profiles in the optical spectra of J1858 also suggests that it has a high velocity ($\sim2500$ km/s) outer accretion disk wind, similar to those observed during the 2015 outburst of V404 Cyg \citep{2016Natur.534...75M,2019ATel12881....1M}. Lastly, low frequency QPOs have been detected in both V404 Cyg and J1858 \citep{2017ApJ...834...90H,2019ATel12512....1H}.

V4641 Sgr had a major outburst in 1999, which showed short duration large amplitude flares similar to those observed in V404 Cyg and J1858 \citep{2000ApJ...528L..93W,2002A&A...391.1013R}. Additionally, it was suggested that an enshrouding envelope, which caused variable absorption, surrounded the inner accretion disk of V4641 Sgr \citep{2002A&A...385..904R}. This source has also exhibited a fast wind, with velocities up to $\sim3,000$ km/s, and also showed P-Cygni profiles in its optical spectra, suggesting that it is coming from the outer accretion disk \citep{2005MNRAS.363..882L,2018MNRAS.479.3987M}.

While both V404 Cyg and V4641 Sgr are similar to J1858 in many ways, there still remain two distinct differences between the first two systems and J1858. The first difference is that the orbital inclination angles for both V404 Cyg and V4641 Sgr are relatively large, $\sim67^{\circ}$ and $\sim72^{\circ}$, respectively \citep{2010ApJ...716.1105K,2014ApJ...784....2M}, while the 3$\sigma$ upper-limit on the disk inclination angle for J1858 derived from our reflection fits is relatively  low, $i<29^{\circ}$. These large inclinations in the first two systems imply that the system is being viewed close to edge on, suggesting that the large amounts of variable obscuring material can be explained by a flared disk (possibly with clumps of material) which intersects the line of sight between the observer and the inner accretion disk (see e.g., Figure 8 in \citealt{2017MNRAS.471.1797M}). Given that the inclination of the inner accretion disk of J1858 appears to be low, this explanation appears less likely to be applicable to J1858. However, misalignments of $\sim15^{\circ}$ between the orbit of the system and the inner accretion disk have been observed in Cygnus X-1 (see e.g., \citealt{2014ApJ...780...78T,2016ApJ...826...87W}) and even larger misalignments (possibly $\sim30^{\circ}-50^{\circ}$) have been observed in the two systems V4641 Sgr and V404 Cyg \citep{2002MNRAS.336.1371M,2019Natur.569..374M}, which are very similar to J1858. Thus, the possibility of a large misalignment between the inclination of the orbit and inner accretion disk, leading to the obscuration of the inner regions of the accretion disk by a flared disk, cannot be entirely ruled out. 

An alternative possibility is that there may be some systematic effects on the derived inclination of the inner accretion disk if there is a complicated source geometry (e.g., a thick disk; \citealt{2018ApJ...855..120T}) near the inner accretion flow, which is currently unaccounted for in the simplified RELXILL reflection models. For instance, using a reflection model similar to the one used here (i.e., {\tt relxill\_lp+xillver}), \cite{2017ApJ...839..110W} found  a range of inclinations for V404 Cyg, spanning $i=27^{\circ}-52^{\circ}$. Furthermore, \cite{2019arXiv190712114T} found an inclination of the inner accretion disk ($\sim40^{\circ}$) that also largely differed from the well-determined binary inclination ($\sim 75^{\circ}$) in XTE J1550$-$564.  We also note that lamp post geometry is an idealized, point source geometry, so if the corona is vertically or horizontally extended it may also impact the inferred inclination. Additionally, if the illuminating X-ray source is associated with the base of the jet, it may be moving with a mildly relativistic velocity, which could also affect the inferred inclination (see e.g., Section 5.3 in \citealt{2019arXiv190712114T}.) In any case, since the binary inclination for J1858 still remains unknown, the disk inclination cannot be compared to the binary inclination to look for a possible misalignment or discrepancy. However, since it appears that the source was detected in optical by Pan-STARRs prior to its outburst, we reiterate that it may be possible to constrain the binary inclination once the source returns to quiescence. 

The second way in which J1858 differs from V404 Cyg and V4641 Sgr, is that it appears to be accreting at only a few percent of the Eddington luminosity, whereas V404 Cyg and V4641 Sgr were accreting at, or possibly even above, the Eddington limit during their outbursts \citep{2002A&A...391.1013R,2017MNRAS.471.1797M}. The distance is not well known for J1858, so once better constraints are placed on its distance a clearer picture of its accretion rate will emerge. However, even if the distance to J1858 is as large as 10 kpc, the source luminosity would still only be $\sim10-20\%$ of the Eddington luminosity. One other possible explanation for this difference in luminosities is that for large enough absorbing column densities (i.e., $N_{\rm H}\gtrsim10^{24}$ cm$^{-2}$), scattering processes become important. For example, during one of V404 Cyg's plateau states, \cite{2017MNRAS.468..981M} found a luminosity of only a few percent of the Eddington luminosity when using a reflection model having a large best-fit absorbing column density (i.e., $N_{\rm H}\approx1-3\times10^{24}$ cm$^{-2}$), but after using a model that also accounted for scattering, they found an increase in intrinsic flux by a factor of $\sim30$, pushing the true luminosity to the Eddington luminosity. This explanation is also somewhat  unlikely though, considering that during flaring episodes (e.g., during flares 1 and 2), J1858's intrinsic absorption is about a factor of 20 lower than in V404 Cyg during this plateau state.

\section{Summary}
\label{summ}

We have reported on the first {\sl NuSTAR} observation of Swift J1858.6$-$0814, which is a newly discovered BH binary candidate. The main findings of this work can be summarized as follows:\\
$\bullet$The source exhibits large amplitude flares, showing an increase in count rate by a factor of $\sim100$ on timescales of $10-100$ s, but we find no evidence of any periodicity or QPOs in the {\sl NuSTAR} light curves. \\
$\bullet$The observed flares are accompanied by large changes in the source's hardness ratio, suggesting that the source's spectrum also significantly changes during the flares. We split the source's energy spectra into three different modes based on where the source is located in the HID during a given interval. We then fit these spectra with a relativistic reflection model, allowing us to constrain a number of the source's physical parameters. \\
$\bullet$Interestingly, we find a large and variable partially covering absorbing column density ($N_{\rm H}=14-42\times10^{22}$ cm$^{-2}$) dependent on the mode of the source. We also find that a thermal component is required to adequately fit the spectra. These results are supported by the joint fits to the energy spectra from two flares simultaneously observed by {\sl NuSTAR} and {\sl NICER}. \\
$\bullet$We constrain the BH spin to be $a^{*}>0.0$ at the 90\% level and $a^{*}>-0.8$ at the 3$\sigma$ level, assuming the inner radius of the accretion disk is at the ISCO. \\
$\bullet$The inclination of the inner accretion disk derived from our fits appears to be relatively low ($i<29^{\circ}$ 3$\sigma$ upper-limit), making the origin of the large amount of obscuring material unclear. Future comparisons of the derived inner accretion disk inclination with the binary inclination angle (once it is known) can help further our understanding of this system. \\
$\bullet$The source shows many similarities to the well-known Galactic BH binaries V404 Cyg and V4641 Sgr. However, J1858's low inclination and low luminosity differ greatly from these two sources, making a direct comparison difficult at this point in time. \\
This source was followed-up by a large multiwavelength campaign, including five additional {\sl NuSTAR} observations. Therefore our understanding of this interesting source will continue to grow as more results are released. 

\software{XSPEC (v12.10.1; \citealt{1996ASPC..101...17A}), NUSTARDAS (v1.8.0), NICERDAS (V005), Stingray \citep{2017ApJ...834...90H}, Xselect (v2.4e), Matplotlib \citep{2007CSE.....9...90H}, HEASOFT (v6.25), MWDust \citep{2016ApJ...818..130B}}

\medskip\noindent{\bf Acknowledgments:} This work made use of data from the {\it NuSTAR} mission, a project led by the California Institute of Technology, managed by the Jet Propulsion Laboratory, and funded by the National Aeronautics and Space Administration. We thank the {\it NuSTAR} Operations, Software and  Calibration teams for support with the execution and analysis of these observations.  This research has made use of the {\it NuSTAR}  Data Analysis Software (NuSTARDAS) jointly developed by the ASI Science Data Center (ASDC, Italy) and the California Institute of Technology (USA).  JH and JAT acknowledge partial support from {\em NuSTAR} Guest Observer grant 80NSSC19K0404. JH acknowledges support from an appointment to the NASA Postdoctoral Program at the Goddard Space Flight Center, administered by the USRA through a contract with NASA. DJW acknowledges support from STFC in the form of an Ernest Rutherford Fellowship. MC acknowledges support from the Centre National d`Etudes Spatiales (CNES).  We thank the anonymous referee for providing useful comments which have improved the  quality of this paper.

\end{document}